\documentclass[aps,pra,floatfix,superscriptaddress,showpacs,twocolumn,nofootinbib,longbibliography]{revtex4-1}

\usepackage{endnotes}
\usepackage{amsfonts}
\usepackage{amsmath}
\usepackage{amssymb}
\usepackage{mathtools}
\usepackage{graphicx}
\usepackage{color}
\usepackage{changes}
\usepackage{bbm}
\usepackage[hidelinks]{hyperref}
\hypersetup{
     colorlinks   = true,
     citecolor    = blue,
     linkcolor    = blue
}
\usepackage{matlab-prettifier}
\usepackage{ulem}

\newcommand{\ignore}[1]{}
\newcommand{\nobibentry}[1]{{\let\nocite\ignore\bibentry{#1}}}

\newcommand{\bibfnamefont}[1]{#1}
\newcommand{\bibnamefont}[1]{#1}
\newcommand{\average}[1]{\left<#1\right>}
\newcommand{\ket}[1]{\left\vert#1\right\rangle}
\newcommand{\bra}[1]{\left\langle#1\right\vert}

\begin{document}
\title{Bath-induced correlations enhance  thermometry precision at low temperatures}

\author{Guim Planella}
\address{ICFO-Institut de Ciencies Fotoniques, The Barcelona Institute of
	Science and Technology, 08860 Castelldefels (Barcelona), Spain}
\address{Facultat de F\'isica, Universitat de Barcelona, 08028 Barcelona, Spain}
\address{Institute for Theoretical Physics, Utrecht University, 3584 CS Utrecht, Netherlands}
\author{Marina F. B. Cenni}
\address{ICFO-Institut de Ciencies Fotoniques, The Barcelona Institute of
	Science and Technology, 08860 Castelldefels (Barcelona), Spain}
\author{Antonio Ac\'in}
\address{ICFO-Institut de Ciencies Fotoniques, The Barcelona Institute of
Science and Technology, 08860 Castelldefels (Barcelona), Spain}
\address{ICREA-Instituci\'o Catalana de Recerca i Estudis Avan\c cats, 08010, Barcelona, Spain}
\author{Mohammad Mehboudi}
\address{ICFO-Institut de Ciencies Fotoniques, The Barcelona Institute of Science and Technology, 08860 Castelldefels (Barcelona), Spain}
\address{Max-Planck-Institut f\"ur Quantenoptik, D-85748 Garching, Germany}
\address{D\'epartement de Physique Appliqu\'ee, Universit\'e de Gen\`eve, 1211 Gen\`eve, Switzerland}
\begin{abstract}
We study the role of bath-induced correlations in temperature estimation of cold Bosonic baths.
Our protocol includes multiple probes, that are not interacting, nor are they initially correlated to each other. 
They interact with a Bosonic sample and reach a non-thermal steady state, which is measured to estimate the temperature of the sample.
It is well-known that in the steady state such non-interacting probes may get correlated to each other and even entangled. Nonetheless, the impact of these correlations in metrology has not been deeply investigated yet. Here, we examine their role for thermometry of cold Bosonic gases and show that, although being classical, bath-induced correlations can lead to significant enhancement of precision for thermometry. The improvement is especially important at low temperatures, where attaining high precision thermometry is particularly demanding. 
The proposed thermometry scheme does not require any precise dynamical control of the probes and tuning the parameters and is robust to noise in initial preparation, as it is built upon the steady state generated by the natural dissipative dynamics of the system. 
Therefore, our results put forward new possibilities in thermometry at low temperatures, of relevance for instance in cold gases and Bose--Einstein condensates.
\end{abstract}
\maketitle

{\it Introduction.---}Achieving
extremely low temperatures is a must for quantum simulation and computation in many platforms.
In order to fully characterize any system that works for such tasks, aside from tunable parameters one has to estimate the non-tunable ones as well. Although these parameters vary depending on the platform, temperature is common among almost all, because thermal states naturally appear in many physical systems. Even if that is not the case, the statistics of sub-systems of a quantum system often behave \textit{as if} the quantum system was at thermal equilibrium~\cite{Gogolin_2016,Popescu2006,PhysRevLett.96.050403}. Therefore, thermometry is a major focus of many theoretical and experimental research carried out in quantum systems~\cite{Mehboudi_2019,de2018quantum,PhysRevLett.114.220405,de2016local,Paris_2015,PhysRevLett.122.030403,razavian2019quantum,PhysRevB.98.045101,Mehboudi_2015,PhysRevA.103.023317,PhysRevLett.119.090603,PhysRevLett.123.180602,PhysRevA.95.022121,Campbell_2017,PhysRevLett.125.080402,mok2021optimal,Latune_2020,rubio2020global,Potts2019fundamentallimits,PhysRevResearch.2.033394}.

Since quantum systems, especially when made of many constituents, are fragile and costly to prepare, the usage of small systems as \textit{probes} is an essential method for non-destructively measuring their parameters~\cite{Giovannetti2011}. As such, individual quantum probes for thermometry have been studied in several scenarios~\cite{PhysRevLett.122.030403,mukherjee2019enhanced,PhysRevA.96.062103} and their usefulness was recently demonstrated experimentally in ultracold gases~\cite{bouton2019single}. When the probe thermalizes with the sample, universal results can be obtained thanks to the Gibbs ensemble, that connects thermometry precision to the heat capacity~\cite{PhysRevLett.114.220405,Paris_2015}. At very low temperatures, however, quantum probes do not thermalize with the sample; they rather reach a non-thermal steady state (NTSS)\footnote{Here, we refer to the reduced state of the probe as non-thermal steady state (NTSS) to emphasise that it cannot be written as a Gibbs state with the Hamiltonian $H_{\rm p}$. To avoid confusion, we do not use the terminology non-equilibrium steady state (NESS), because unlike the latter, here the total system-Bath might be described by a Gibbs state---roughly speaking, the set of NTSSs contains NESSs. See e.g., \cite{aschbacher2006topics,PhysRevE.100.022105,bach2000return} for further details.}, which is generally model-dependent~\cite{breuer2002theory,PhysRevA.96.062103}. 
Thermometry at low temperatures is timely and challenging. It has been shown that for thermal equilibrium probes, the error diverges exponentially as $T\to 0$ \cite{Paris_2015,PhysRevLett.114.220405}. Thus many attempts have been addressed to overcome this limitation. In particular, when the probes are not at thermal equilibrium, it was shown that instead of the exponential one can achieve a polynomial divergence \cite{PhysRevA.96.062103,PhysRevResearch.2.033394,PhysRevB.98.045101,Potts2019fundamentallimits}; still the error can be very large. Our work is thus motivated by these challenges, specifically regarding experimentally relevant bosonic models that describe impurity based thermometry of ultra-cold gases.

Most of major experimental thermometry protocols that address ultracold gases use the time-of-flight absorption technique, which can be very precise, but is often destructive~\cite{leanhardt2003cooling,PhysRevLett.96.130404,Gati_2006}. Nonetheless, there are some experiments in which an impurity is used as a probe. This impurity can be made up of \textit{multiple} atoms that simultaneously interact with the system. They have been extensively studied and experimentally realised in both bosonic and fermionic gases~\cite{PhysRevLett.103.223203,olf2015thermometry,Massignan_2014,PhysRevLett.117.055302,PhysRevLett.103.223203,PhysRevLett.89.093001,PhysRevX.6.041041,PhysRevA.85.023623,PhysRevA.85.021602,PhysRevLett.117.055301,mukherjee2019enhanced,PhysRevA.71.023606,PhysRevLett.96.210401,PhysRevLett.109.235301,PhysRevA.88.053632,PhysRevA.89.033615,PhysRevLett.115.160401,PhysRevLett.115.125302,PhysRevLett.117.113002}.

It is well known that a quantum bath/sample can create correlations, and sometimes entanglement, among sub systems of a probe that interact with it. This can be the case even if the probes are initially uncorrelated and/or if they do not interact directly with one another. We call this phenomenon \textit{bath-induced correlations}. 
In the past few years, several theoretical works have reported bath-induced correlations/entanglement at NETS of different platforms including Bosonic and Fermionic environments~\cite{Wolf_2011,PhysRevA.88.042303,PhysRevA.78.042307,PhysRevA.82.012333,PhysRevE.91.062123,10.21468/SciPostPhys.6.1.010} and even realized them experimentally~\cite{PhysRevLett.107.080503}. However, to our knowledge, the use of such correlations to estimate the bath parameters has not been studied.

The main goal of our work is to analyze the thermometry of bosonic systems when using multiparticle probes and investigate the impact of bath-induced correlations in precision thermometry. We show that although entanglement---quantified by negativity---is absent, yet bath-induced correlations help significantly improving the precision of thermometry and in some cases leading to two orders of magnitude increase in the quantum Fisher information. 
These results can be used to address and improve non-demolition thermometry of Bose--Einstein condensates (BECs) in the nK and sub-nK domain aligned with previous efforts in characterizing correlations in BECs~\cite{10.21468/SciPostPhys.6.1.010}.

\begin{figure}
	\includegraphics[width=1\linewidth]{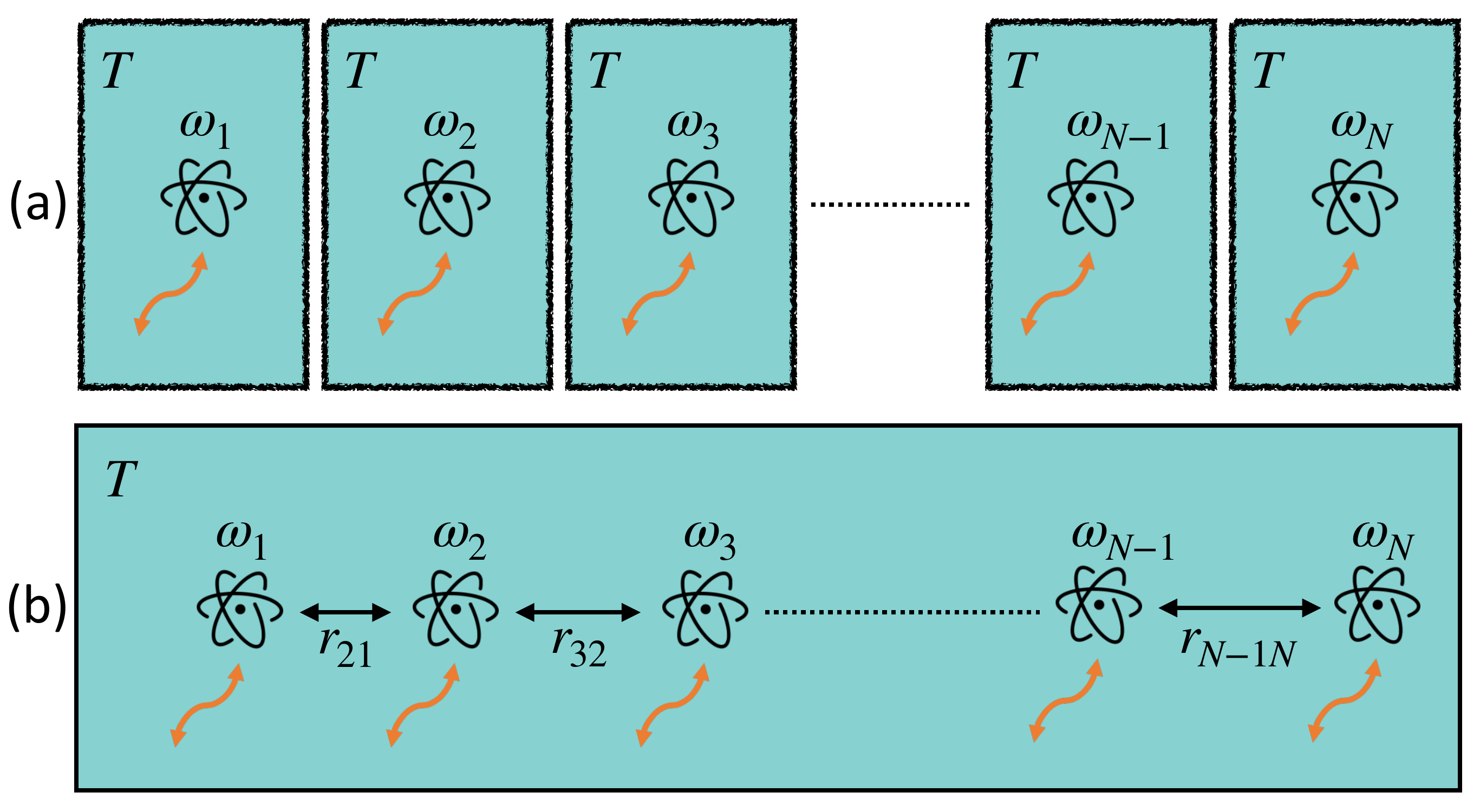}
	\caption{Schematic of the thermometry protocols. (a) Independent baths: each thermometer is in contact with a separate bath, thus no correlations are built among them. Hence, the precision is additive. (b) Common bath---focus of this work: all thermometers are in contact with the same bath. The $i$th and the $j$th thermometers have a distance $r_{ij}$, they \textit{do not interact} with each other nor do they share initial correlations, yet they get correlated thanks to their interactions with the common bath. Our results show that such correlations might lead to super-additive precision at low temperatures. The two scenarios are equivalent when the distance between the probes is very large or at high temperatures.}
	\label{fig:Setting}
\end{figure}

{\it The setup and the model.---}We consider a sample (bath) of Bosonic harmonic oscillators. It is in a thermal state, and our aim is to estimate its temperature $T$ by bringing it in contact with an external probe. After sufficiently long interaction among the probe and the sample, the probe relaxes to its NTSS. 
Measurements are carried out solely on the probe, hence realizing a non-demolition measurement on the sample. 
Figure \ref{fig:Setting} illustrates the scenarios that we address here: (a) the independent-bath scenario, where no correlations are created among different oscillators as they lie in separate parts of the bath~\footnote{More precisely, we define this scenario as the case where the  inter-probe distances are larger than the correlation length of the bath.}. This is our reference scenario. Any situation in which one invokes the thermalization assumption can be analyzed within this framework. (b) a more realistic scenario, where all of the probes are embedded in the same bath. If we increase the distance among neighboring probes, we expect to revive (a). Otherwise, this scenario gives rise to correlations among different oscillators---see bellow. This implies that using (a) to describe realistic protocols can lead to significant miscalculation of the thermometry precision. Moreover, while the thermometry precision in the independent-bath scenario (a) is additive, the correlations that appear in the common bath scenario (b) can lead to super additive precision for thermometry giving rise to a significant enhancement. 

Let us introduce details of the model and sketch how to exactly solve for the probe's NTSS (more details are provided in the Appendix, see also \cite{weiss2012quantum,BreuerBook,PhysRevA.96.062103,weedbrook2012gaussian}). 
The total Hamiltonian reads
\begin{align}\label{eq:Ham_total}
	H = H_{\rm p} + H_{\rm B} + H_{{\rm PB}}.
\end{align}
where the Hamiltonian of the $1d$ probe (see \cite{PhysRevE.91.062123} for extension to higher dimensions) is
\begin{align}\label{eq:Ham_probe}
	H_{\rm P} = \sum_{i=1}^{N} \frac{p_i^2}{2m_i} + \frac{1}{2}\sum_{i=1}^N m_i\omega_i^2 x_i^2 + \frac{1}{2}\sum_{i\neq j}^{N} g_{i j} x_ix_j,
\end{align}
with $x_i$ being the displacement from \textit{equilibrium position} $r_i$ of the $i$th probe oscillator and $p_i$ being the conjugate momentum. For now we allow for inter-oscillator interactions with couplings $g_{ij}$, but to study only bath-induced correlations we set $g_{i j}=0$ in simulations. 
The Hamiltonian of the bath reads
\begin{align}\label{eq:Ham_Bath}
	H_{\rm B} = \sum_{k} \frac{q_{k}^2}{2m_{k}} + \frac{1}{2}\sum_{k} m_{k}\omega_{k}^2 y_{k}^2,
\end{align}
being $y_{k}$ and $q_{k}$ the position and momentum of the bath mode with the wave vector ${{k}}$.
Generally, $H_{\rm B}$ could be an interacting model, however, if the interaction is quadratic, one can always bring it to the form \eqref{eq:Ham_Bath} by finding its normal modes (e.g., see the $1d$ BECs studied in~\cite{Lampo2017bosepolaronas,PhysRevLett.122.030403}). 
Finally, we consider a probe--bath interaction of the form
\begin{align}
	H_{\rm PB} = \sum_{i=1}^N\sum_{k}G_{k}x_i\left( y_{k} \cos{{k}  {r}_{i}}+\frac{q_{k}}{m_{k}\omega_{k}} \sin{{k} {r}_{i}}\right),
\end{align}
which is valid under the \textit{long wave approximation}, 
where the wavelength of the bath excitations is much larger that the displacement of the oscillators from equilibrium i.e., $|k x_i|\ll 1$ ~(see e.g., \cite{PhysRevA.88.042303}).

Equation \eqref{eq:Ham_total} is quadratic, thus the dynamics is Gaussian and the NTSS will be Gaussian too. The NTSS does not depend on the initial state of the probe, it only depends on the total Hamiltonian, as well as the initial state of the bath, which we choose to be thermal at (unknown) temperature $T$. Since the NTSS is Gaussian, we only need the first and second order correlations---that is, the \textit{displacement vector} and the \textit{covariance matrix}, respectively---to fully describe it. If we define $R = (x_1,p_1,\dots,x_N,p_N)^T$, then the displacement vector is a $2N$ dimensional vector with elements $d_i = \average{R_i}$ while the covariance matrix is a $2N\times 2N$ symmetric matrix with elements $\Gamma_{ij} = \average{\{R_i, R_j\}}/2 - \average{R_i}\average{R_j}$.

The conventional method of finding $d$ and $\Gamma$ starts by using the Heisenberg equations of motion---that for any observable $O$ reads ${\dot O} = i[H,O]$.
Applying to all degrees of freedom gives
\begin{align}\label{eq:x_dot}
	{\dot x}_{i} & =  \frac{p_i}{m_i},\\
	\label{eq:y_dot}
	{\dot y}_{k} & = \frac{q_{k}}{m_{k}}+\sum_{i=1}^{N }\frac{G_{k}}{m_{k}\omega_{k}} x_i \sin {{k}}  {r}_i,\\
	\label{eq:p_dot}
	{\dot p}_{i} & = -m_i\omega_i^2x_i-\sum_{j\neq i}g_{ij}x_j \nonumber\\ &-\sum_{k}G_{k}\left( y_{k}\cos {{k}}  {r}_i + \frac{q_{k}}{m_{k} \omega_{k}}\sin {k}  {r}_i \right) ,\\
	\label{eq:q_dot}
	{\dot q}_{k} & = - m_{k}\omega_{k}^2y_{k} - \sum_{i=1}^{N}G_{k}x_i \cos {k} {r}_i.
\end{align}
Solving these equations for the probe degrees of freedom, gives the quantum Langevin equations of motion (see the Appendix and refs.~\cite{Langevin1908,BreuerBook})
\begin{align}\label{eq:QLE_time}
	m_i {\ddot x}_i +m_i\omega_i^2 x_i + \sum_{j=1}^Ng_{ij}x_j - \sum_{j=1}^N \chi_{ij}\star x_j = F_i,
\end{align}
where $\star$ stands for convolution. Here, the susceptibility matrix reads
\begin{align}\label{eq:def_chi}
	\chi_{ij}(t) = \sum_{k} \frac{G^2_{k}}{m_{k}\omega_{k}}\sin(\omega_{k}t + {k}   {r}_{ij} )\Theta(t),
\end{align}
where we define $ {r}_{ij}\coloneqq {r}_i - {r}_j$, and the step function $\Theta(t)$ imposes causality. The susceptibility matrix is responsible for both memory effects---through the convolution---and correlations among the probes, even if $g_{ij}=0$.
Finally, the vector of Brownian forces reads
\begin{align}
	F_i(t)=&-\sum_{k} G_{k}\big( y_{k}\left( t_0 \right) \cos\left( \omega_{k}\left( t-t_0 \right)+{k} {r}_i \right)\nonumber\\
	&+\frac{q_{k}\left( t_0 \right) }{m_{k}\omega_{k}}\sin\left( \omega_{k}\left( t-t_0 \right)+{k} {r}_i \right) \big).
\end{align}
The solution of~\eqref{eq:QLE_time} depends on the probe-bath interaction, and the particular spectral density describing it. The latter is a matrix with the elements
\begin{align}\label{eq:def_SD}
J_{ij}(\omega) &= \sum_{k}\frac{\pi G^2_{k}}{2m_{k}\omega_{k}}\cos (  {k}   {r}_{ij} )\delta(\omega - \omega_{k}).
\end{align}
In what follows we consider a $1d$ Bosonic bath with linear dispersion ${k}=\omega_{k}/c$, where $c$ denotes speed of sound. This corresponds to an Ohmic spectral density \cite{gonzalez2017testing,kleinekathofer2004non}
\begin{align}\label{eq:SD_Freq}
	J_{ij}(\omega)={\gamma^2} \omega \frac{\Omega^2}{\omega^2+ \Omega^2}\cos \frac{\omega \left|{r}_{ij}\right|}{c},
\end{align}
with $\gamma$ the probe-Bath interaction strength and $\Omega$ the cutoff frequency.
We exactly solve~\eqref{eq:QLE_time} and characterize the steady state by finding $d$ and $\Gamma$. Firstly, we find that 
$d=0$. Secondly, if $\Gamma^{(\rm a)}(T)$ and $\Gamma^{(\rm b)}(T)$ are the \textit{temperature dependent} covariance matrices in scenarios (a) and (b), respectively, we observe a major difference in their correlations: While in (a) we have an uncorrelated state with $\Gamma^{(\rm a)}(T) = \oplus_i \sigma_{i}^{(\rm a)}(T)$---$\sigma_{i}^{(\rm \alpha)}(T)$ being the local covariance matrix of the $i$th probe in scenario $\alpha \in \{\rm a,b\}$---this is not the case for (b), in which inter-oscillator correlations appear.
These correlations disappear at large distance, and/or at high temperatures. Nonetheless, at low temperature regimes, and small distances they can be significant---see the Appendix for details---and are responsible for enhanced thermometry precision.
Interestingly, we find out that entanglement negativity, for arbitrary bipartitions, is zero. Nonetheless, classical correlations are present in the common bath scenario and
we show below that this major difference in the correlation structure significantly enhances thermometry precision.

{\it Metrology in Gaussian quantum systems.---}We
are dealing with parameter estimation in Bosonic Gaussian quantum systems. Let $\Gamma(\lambda)$ be the covariance matrix of a Gaussian quantum system. Here, $\lambda$ is to be estimated, which can be temperature or any other parameter. We drop the parameter-dependence of the covariance matrix to have a lighter notation, and restrict ourselves to scenarios with $d=0$, as is the case in our problem.

For a given measurement, with the measurement operator set $\{\Pi^s(o)\}$---where $s$ labels the specific measurement, and $o$ denotes different outcomes which can be continuous or discrete---the error on estimation of $\lambda$ is bounded from below by~\cite{doi:10.1142/S0219749909004839}
\begin{align}\label{eq:CRB_QCRB}
\delta \lambda(s) \geq \frac{1}{\nu \sqrt{{\cal F}^{\rm cl}(\lambda,s)}}\geq\frac{1}{\nu \sqrt{{\cal F}^{\rm Q}(\lambda)}}\eqqcolon \delta \lambda_{\min}
\end{align}
where $\nu$ is the number of measurements, and ${\cal F}^{\rm cl}(\lambda,s)$ is the classical Fisher information (CFI) associated with the performed measurement defined as
\begin{align}\label{eq:cfi}
{\cal F}^{\rm cl}(\lambda,s) \coloneqq \average{\left[\partial_{\lambda} \log p(o|s,\lambda)\right]^2}_{p(o|s,\lambda)}.
\end{align}
Here, $p(o|s,\lambda)$ is the conditional probability of observing $o$ given the parameter has the value $\lambda$ and the measurement $s$ is performed.
The quantity ${\cal F}^{\rm Q}(\lambda) \coloneqq {\max_s}~{\cal F}^{\rm cl}(\lambda,s)$ is the quantum Fisher information (QFI) that is obtained by maximizing the CFI over all measurements and thus is independent of $s$.
The first inequality in \eqref{eq:CRB_QCRB} is called the Cramer-Rao bound, which can be saturated by suitable post-processing of the outcomes. Thus, CFI can be a precision quantifier for a given measurement. The second inequality is the quantum Cramer-Rao bound that sets a fundamental lower bound on the error, regardless of the measurement. Importantly, this bound can be also saturated, by definition of the QFI. 

\begin{figure}[t!]
	\includegraphics[width=1\linewidth]{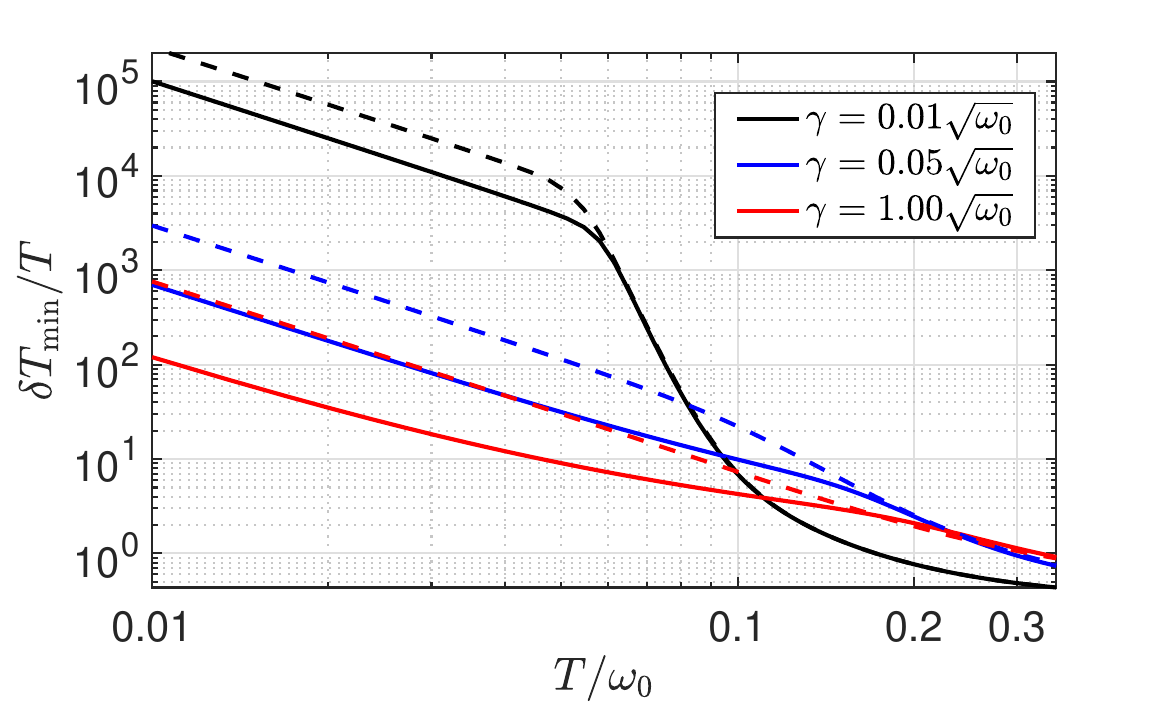}
	\caption{Minimum relative error ($\delta T_{\min}/T$ from r.h.s of Eq.~\eqref{eq:CRB_QCRB}) against temperature for different system-bath couplings $\gamma$ (quantitative description in the Appendix) for a fixed number of oscillators. The solid curves represent the common bath scenario (b) whereas the dashed lines are obtained considering independent baths scenario (a). We see that at low temperatures by increasing $\gamma$ the error reduces, whereas at higher temperatures the opposite holds (not shown). Moreover, embedding the oscillators at the same bath can significantly decrease the relative error.
	Here, we set the parameters to $\omega_0=1$ in arbitrary units, $\omega_i=\omega_0\forall i$, $r_{21}/c\omega_0=0.1$, $\Omega=100~\omega_0$, $g_{ij}=0$, and $N=10$. }\label{fig:1}
\end{figure}

\begin{figure}
	\includegraphics[width=1\linewidth]{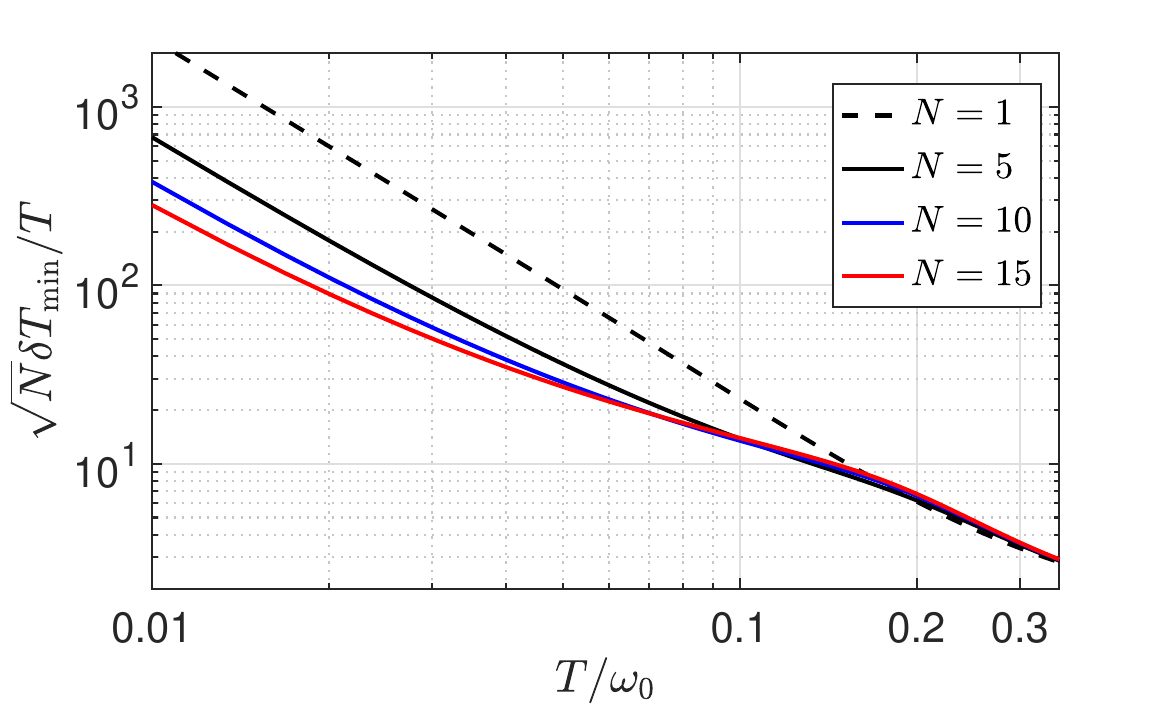}
	\caption{Minimum relative error normalized by the number of oscillators (that is $\sqrt{N}\delta T_{\min}/T$), as a function of temperature, in the independent baths scenario (a) for arbitrary $N$ (dashed black) and in the common bath scenario (b): for different number of oscillators  $N=1$ (also dashed black), $N=5$ (solid black), $N=10$ (solid blue) and $N=15$ (solid red). Unlike the scenario (a) which is additive, the scenario (b) is super additive, as the normalised error reduces by increasing $N$.
	The parameters are set to $\omega_0=1$ in arbitrary units, and $\omega_i=\omega_0\forall i$, $r_{21}/c\omega_0=0.1$, $\Omega=100~\omega_0$, $g_{ij}=0$, and $\gamma=\sqrt{\omega_0}$.}\label{fig:2}
\end{figure}

Finding the optimal measurement and the QFI is challenging and requires different approaches depending on the platform, the underlying dynamics, and the specific parameter to be estimated.
Nonetheless, for Gaussian systems, one can routinely find them~\cite{monras2013phase,PhysRevA.89.032128,PhysRevA.98.012114,malago2015information,Nina_et_al_in_preparation}. In the Appendix we show that the optimal measurement is given by a linear combination of second order quadratures, which is highly non-local and experimentally demanding. Thus we also examine practically feasible alternatives, namely local position and momentum measurements defined as $X = \otimes_{i=1}^N x_i$ and $P = \otimes_{i=1}^N p_i$, respectively. These belong to the family of Gaussian measurements---i.e., when measuring Gaussian systems their outcomes are Gaussian distributed---for which the CFI is straightforwardly calculable~\cite{monras2013phase,Nina_et_al_in_preparation}. 
Despite being sub-optimal measurements, they can benefit from the enhancement, which puts forward an experimentally realisable scheme to significantly enhancing thermometry of Bosonic systems.
\noindent
{\it Enhanced thermometry with bath-induced correlations.---}We
study the single shot ($\nu=1$) relative error $\delta T_{\rm min}/T$ 
for a variety of parameters.
Figure~\ref{fig:1} illustrates $\delta T_{\rm \min}/T$ versus temperature for various probe-bath couplings. For both scenarios (a) and (b), we observe that at high temperatures $\delta T_{\rm min}/T$ increases by increasing the coupling, whereas at low temperatures the opposite is observed. This unifies the findings of \cite{PhysRevLett.122.030403} and \cite{PhysRevA.96.062103}, and extends them from the single to the multi-probe scenario. Furthermore, Fig.~\ref{fig:1} shows that at lower temperatures and for a fixed coupling, the scenario (b) always outperforms (a).
To see this more clearly, we fix the coupling, and depict the normalised error $\sqrt{N}\delta T_{\rm min}/T$ in Fig.~\ref{fig:2}. At small temperatures, we observe that in scenario (b) the normalized error reduces by increasing the number of probes and thus significantly outperforming scenario (a). By increasing $T$, the enhancement is progressively lost and even the scenario (a) might become slightly better. At even higher temperatures, the probes \textit{thermalize} with the bath, and the two scenarios become equivalent as expected.


\begin{figure}
	\includegraphics[width=1\linewidth]{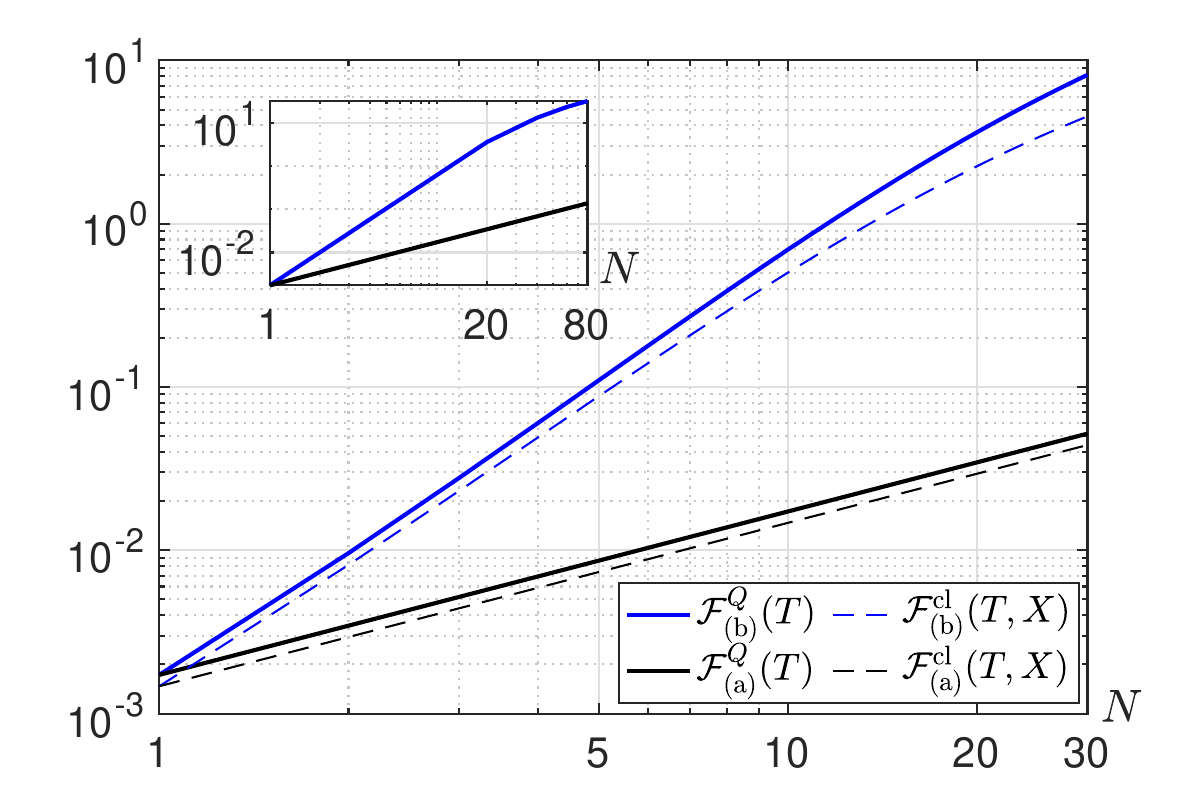}
	\caption{Loglog plot of the quantum Fisher information vs $N$ in solid blue for scenario (b) and solid black for scenario (a). Here, we tune the setting to the \textit{low temperature} limit with $T_0=0.01 \omega_0$. Scenario (b) behaves super-linearly; our data-fitting shows that ${\cal F}(T_0) \propto N^{2.5}$ in the range of $1<N<30$, although the slope is decreasing as we see for $N\approx 80$ we have ${\cal F}(T_0) \propto N^{1.1}$ (Inset). Nonetheless, for $N=30$ we already see that the QFI in scenario (b) is two orders of magnitude bigger than (a)---which shows a linear behaviour as suggested by the additivity of the QFI for a tensor product state. The CFI for measuring $X = \otimes_{i=1}^N x_i$ is also depicted in dashed black for (a) and dashed blue for (b). Albeit sub-optimal, these local measurements show a similar behavior to the optimal ones in both scenarios. Specifically, for scenario (b) they are super-linear with $N$. 
	The rest of the parameters are set to $\omega_0=1$, $\omega_i=\omega_0\forall i$, $\gamma= \sqrt{\omega_0}$, $\Omega=100 \omega_0$, $g_{ij}=0$ and $r_{21}/c\omega_0=0.1$.}
	\label{fig:scaling_quadratic}
\end{figure}

In Fig.~\ref{fig:scaling_quadratic} we fix the temperature in the low-temperature limit, and 
depict the QFI versus the number of oscillators.
The slope of the QFI in the common bath scenario (b) is clearly larger than scenario (a). For $1<N<30$, on average, ${\cal F}^{Q}(T) \propto N^{2.5}$. However, as seen from the inset, the slope decreases for larger $N$ and it reaches $N^{1.1}$ for $N\approx 80$. Nonetheless, for $N\geq 30$ we see at least \textit{two orders of magnitude} growth in the QFI due to the bath-induced correlations.
It is worth mentioning that in the literature there are few works that study the role of correlations for thermometry in \textit{dynamical scenarios} \cite{PhysRevLett.123.180602,PhysRevA.95.022121} that might need high degrees of control; but to our knowledge this is the first time bath-induced correlations at NTSS are exploited for thermometry---and even other metrological tasks. What is more interesting is that the entanglement, as characterised by logarithmic negativity, is zero among any possible bi-partition, hence, the enhancement is not due to quantum correlations.

Generally, the optimal measurement can be highly non-local. This is indeed the case for our common bath scenario (b)---See the Appendix and Refs. \cite{PhysRevLett.105.020503,adesso2011gaussian,PhysRevA.87.012119}. For experimental purposes we also find simple \textit{local} measurements that exploit the bath-induced correlations for enhanced thermometry at low temperatures. Particularly, we examine the CFI of local position and momentum measurements (only position measurement is shown here.)
The result is depicted in Fig.~\ref{fig:scaling_quadratic}. Although being sub-optimal,
the precision is super additive and is better than the QFI of scenario (a) already for $N\geq 2$, demonstrating the advantage of exploiting bath-induced correlations even with local Gaussian measurements.

{\it Discussion.---}We showed that bath-induced correlations present in the steady state of thermometers can play a prominent role in thermometry of Bosonic baths at \textit{low temperatures}, which is where precision thermometry is known to be most demanding. 
Although these correlations are not of a quantum nature, they can increase the thermometry precision up to two orders of magnitude, for both optimal and experimentally feasible measurements. 
In our scheme, neither preparation of highly entangled states, nor dynamical control is required---compared to alternative dynamical based proposals~\cite{PhysRevLett.125.080402,PhysRevLett.123.180602,PhysRevA.95.022121,mukherjee2019enhanced}---which makes them very appealing.
Unfortunately, unlike typical phase estimation problems, the steady state depends on the temperature non-trivially. This makes it challenging to pinpoint the intrinsic properties of the bath and the steady state that intensify the impact of correlations in thermometry precision. Therefore, an interesting future direction is, to consider properties of covariance matrices and their temperature dependence (regardless of any physical model in the background) that leads to such improvements. It would be also interesting to investigate other spectral densities---corresponding to physically relevant models e.g., with alternative spacial dimension and trapping potential---and see whether/how they affect the impact of the common bath on thermometry precision.
Bath-induced correlations for thermometry might be examined in several experimental setups e.g., \cite{olf2015thermometry,PhysRevLett.85.483,PhysRevA.85.023623,PhysRevA.103.023317,PhysRevLett.111.070401,PhysRevA.89.053617} and possibly explored in fermionic platforms 
\cite{PhysRevLett.102.230402,PhysRevLett.103.223203,PhysRevLett.102.230402,kohstall2012metastability,koschorreck2012attractive}. 

{\it Acknowledgments.---}Constructive
discussions with M.A.~Garcia-March, C.~Charalambous, S.~Iblisdir, S.~H\'ernandez-Santana, D.~Alonso and L.A.~Correa are appreciated. This work was financially supported by Spanish MINECO (ConTrAct FIS2017-83709-R, FIS2020-TRANQI and Severo Ochoa CEX2019-000910-S), the ERC AdG CERQUTE, the AXA Chair in Quantum Information Science, Fundacio Cellex, Fundacio Mir-Puig, Generalitat de Catalunya (CERCA, AGAUR SGR 1381 and QuantumCAT) and the Swiss National Science Foundations (NCCR SwissMAP).

\bibliography{Refs}
\appendix
\pagebreak
\onecolumngrid
\section{The quantum Langevin equations of motion and the steady state}\label{App:QLEs}
Starting from equations (5-8) of the main text we can find the quantum Langevin equation by solving for the degrees of freedom of the bath. It is more convenient to  define the creation and annihilation operators 
\begin{align}
	a_{ k}=\sqrt{\frac{m_{ k}\omega_{ k}}{2}} \left( y_{ k} +  \frac{i }{m_{ k} \omega_{ k}} q_{ k} \right), \\
	a^\dagger_{ k}=\sqrt{\frac{m_{ k}\omega_{ k}}{2}} \left( y_{ k} -  \frac{i }{m_{ k} \omega_{ k}} q_{ k} \right).
\end{align}
The equations of motion for the $a_k$ and $a_k^{\dagger}$ decouple, yielding
\begin{align}
	{\dot a}_k&=-i \omega_{ k}a_{ k}-\sum_{i=1}^{N}\frac{i G_{ k}e^{i { k} {r }_i } }{\sqrt{2 m_{ k} \omega_{ k}}} \dot{x_i},\\
	{\dot a}^\dagger_{ k}&=i \omega_{ k}a^\dagger_{ k}+\sum_{i=1}^{N }\frac{i G_{ k}e^{-i { k} {r }_i } }{\sqrt{2 m_{{ k}} \omega_{{ k}}}} \dot{x_i}.
\end{align}
These can be solved for any given $x_i(t)$ with solution
\begin{align}
	a_{ k}(t)&=e^{-i \omega_{ k} (t-t_0)}a_{{ k}}(t_0)-i\frac{G_{ k}}{\sqrt{2 m_{{ k}}\omega_{ k}}} \sum_i \int_{t_0}^te^{-i \omega_{ k} (t-s)}e^{i{ k} {r}_i}x_i(s) ds,\\
	a^\dagger_{ k}(t)&=e^{i \omega_{ k} (t-t_0)}a^\dagger_{ k}(t_0)+i \frac{G_{ k}}{\sqrt{2 m_{{ k}}\omega_{ k}}} \sum_i \int_{t_0}^te^{i \omega_{ k} (t-s)}e^{-i{ k} {r}_i}x_i(s) ds.
\end{align}
By transforming back to position and momenta and substituting into the equations of motion for the probe, we get Eq.~(9). Taking the the Fourier transformation from both sides of Eq.~(9) and using the convolution theorem one obtains
\begin{align}
	m_i(\omega_i^2- \omega^2) {\tilde x}_i(\omega) + \sum_{j=1}^Ng_{ij}{\tilde x}_j(\omega) - \sum_{j=1}^N {\tilde \chi}_{ij}(\omega) {\tilde x}_j(\omega) = {\tilde F}_i(\omega),
\end{align}
where the tilde represents the Fourier transform of the original function. In a compact matrix representation, this reads
\begin{align}\label{eq:Frequency_Position_Force}
	{\tilde x}(\omega) = \alpha^{-1}(\omega) {\tilde F} (\omega),
\end{align}
with the matrix $\alpha$ given by
\begin{align}
	\alpha(\omega) = \left[
	\begin{array}{ccc}
		m_1(\omega_1^2 - \omega^2) - {\tilde \chi}_{11}(\omega) & g_{12} - {\tilde \chi}_{12}(\omega) & \dots \\
		g_{21} - {\tilde \chi}_{21}(\omega) & m_2(\omega_2^2 - \omega^2) - {\tilde \chi}_{22}(\omega) & \dots\\
		\vdots& \vdots & \ddots 
	\end{array}
	\right].
\end{align}
To proceed further, we assume an isotropic bath. This means that all quantities only depend on the magnitude of ${ k}$, in particular $\omega_{ k}=\omega_{- k}$. From the definition of $\chi(t)$ given in equation (10) it is clear that $\chi(t)$ is real. Moreover, given any value of $ k$ present in the bath, $- k$ is also present, as waves should be able to propagate in both directions.  Then, taking the transpose of $\chi$ is the same as exchanging the sign of ${r}_{ij}={r}_i- {r}_j$ which doesn't change the final value of $\chi$, this implies that $\alpha(\omega) $ is a symmetric matrix. 

The covariance matrix can be found by first finding the bath correlation functions. Firstly, notice that at the initial time the bath is in a thermal state for which we have $\average{y_{ k}y_{{p}}} = \delta_{{kp}} \coth(\omega_{ k}/2T)/(2m_{ k}\omega_{ k})$, and $\average{q_{ k}q_{{p}}} = \delta_{{kp}}m_{ k}\omega_{ k} \coth(\omega_{ k}/2T)/2$, and $\average{y_{ k}q_{{p}}} = \delta_{{kp}} i/2$. Using these and after a straightforward calculation one can express the bath correlation functions as
\begin{align}\label{eq:Corr_bath}
	\average{F_i(t^\prime)F_j(t^{\prime\prime})}
	&=\sum_{ k}\frac{G^2_{ k}}{2 m_{{k }}\omega_{ k}}
	\big\{
	\coth\frac{\omega_{ k}}{2T}
	\left[
	\cos \left(\omega_{ k}( t^{\prime}-t^{\prime\prime} )\right)\cos ({ k} {r}_{ij})-\sin \left(\omega_{ k}( t^{\prime}-t^{\prime\prime} )\right)\sin ({ k} {r}_{ij})\right]\nonumber\\
	&-i \sin \left(\omega_{ k}( t^{\prime}-t^{\prime\prime} )\right)\cos ({ k} {r}_{ij}) -i\cos \left(\omega_{ k}( t^{\prime}-t^{\prime\prime} )\right)\sin ({ k} {r}_{ij})\big\}
\end{align}
Using once again the isotropy property of the bath, the terms proportional to $\sin ({ k}  {r}_{ij}) $ cancel out of the summation and \eqref{eq:Corr_bath} becomes
\begin{align}
	\average{F_i(t^\prime)F_j(t^{\prime\prime})} = \frac{1}{\pi}\int_{0}^{\infty} J_{ij}(\omega) \left( \cos(\omega(t^{\prime} - t^{\prime\prime} ))\coth(\frac{\omega}{2T}) - i\sin(\omega(t^{\prime} - t^{\prime\prime} )) \right) d\omega,
\end{align}
where $J(\omega)$ is the spectral density defined in (12).
In a compact matrix form we have
\begin{align}
	\average{F(t^\prime)F^T(t^{\prime\prime})} = \frac{1}{\pi}\int_{0}^{\infty} J(\omega) \left( \cos(\omega(t^{\prime} - t^{\prime\prime} ))\coth(\frac{\omega}{2T}) - i\sin(\omega(t^{\prime} - t^{\prime\prime} )) \right) d\omega,
\end{align}
By Fourier transforming the latter result one finds that
\begin{align}
	\average{{\tilde F}(\omega^{\prime}){\tilde F}^T(\omega^{\prime\prime})} = 2\pi\delta(\omega^{\prime}+\omega^{\prime\prime})[\coth(\frac{\omega^{\prime}}{2T}) + 1]\left[J(\omega^{\prime})\theta(\omega^{\prime}) - J(-\omega^{\prime})\theta(-\omega^{\prime})\right].
\end{align}
If we symmetrize this expression, we will have
\begin{align}
	\frac{1}{2}\average{\left\{{\tilde F}(\omega^{\prime}), {\tilde F}(\omega^{\prime\prime})\right\}} = 2\pi\delta(\omega^{\prime}+\omega^{\prime\prime})\coth(\frac{\omega^{\prime}}{2T}) \left[J(\omega^{\prime})\theta(\omega^{\prime}) - J(-\omega^{\prime})\theta(-\omega^{\prime})\right].
\end{align}

With this, one can already calculate the position-position correlations in the frequency domain. For instance, if we are interested in the element $\average{{\tilde x}_i(\omega){\tilde x}_j(\omega)}$ by substituting in \eqref{eq:Frequency_Position_Force} one finds
\begin{align}\label{eq:pos_pos_freq}
	\frac{1}{2}\average{\left\{{\tilde x}_i(\omega^{\prime}), {\tilde x}_j(\omega^{\prime\prime})\right\}} &=\frac{1}{2}\left[ \alpha^{-1}(\omega^{\prime}) \average{\left\{{\tilde F}(\omega^{\prime}), {\tilde F}(\omega^{\prime\prime})\right\}} \alpha^{-1}(\omega^{\prime\prime})\right]_{ij}\nonumber\\
	& = \frac{1}{2}\left[ \alpha^{-1}(\omega^{\prime}) \average{\left\{{\tilde F}(\omega^{\prime}), {\tilde F}(-\omega^{\prime})\right\}} \alpha^{-1}(-\omega^{\prime})\right]_{ij} \delta(\omega^{\prime} + \omega^{\prime\prime}).
\end{align}

In the time domain, we need to find the inverse double Fourier transform of \eqref{eq:pos_pos_freq}, and evaluate at $t=0$. This is
\begin{align}
	\frac{1}{2}\average{\left\{ x_i,x_j \right\}} = \frac{1}{(2\pi)^2} \int_{-\infty}^{\infty} \frac{1}{2}\average{\left\{{\tilde x}_i(\omega), {\tilde x}_j(-\omega)\right\}} d\omega.
\end{align}
Similarly, we can find the position-momentum, and momentum-momentum correlations
\begin{align}
	\frac{1}{2}\average{\left\{ x_i,p_j \right\}} & =  -\frac{im_j}{(2\pi)^2} \int_{-\infty}^{\infty} \frac{1}{2}\omega\average{\left\{{\tilde x}_i(\omega), {\tilde x}_j(-\omega)\right\}} d\omega,\\
	\frac{1}{2}\average{\left\{ p_i,p_j \right\}} &=  \frac{m_im_j}{(2\pi)^2} \int_{-\infty}^{\infty} \frac{1}{2}\omega^2\average{\left\{{\tilde x}_i(\omega), {\tilde x}_j(-\omega)\right\}} d\omega.
\end{align}

\begin{figure}[t!]
	\begin{tabular}{c}
		\includegraphics[width=.5\columnwidth]{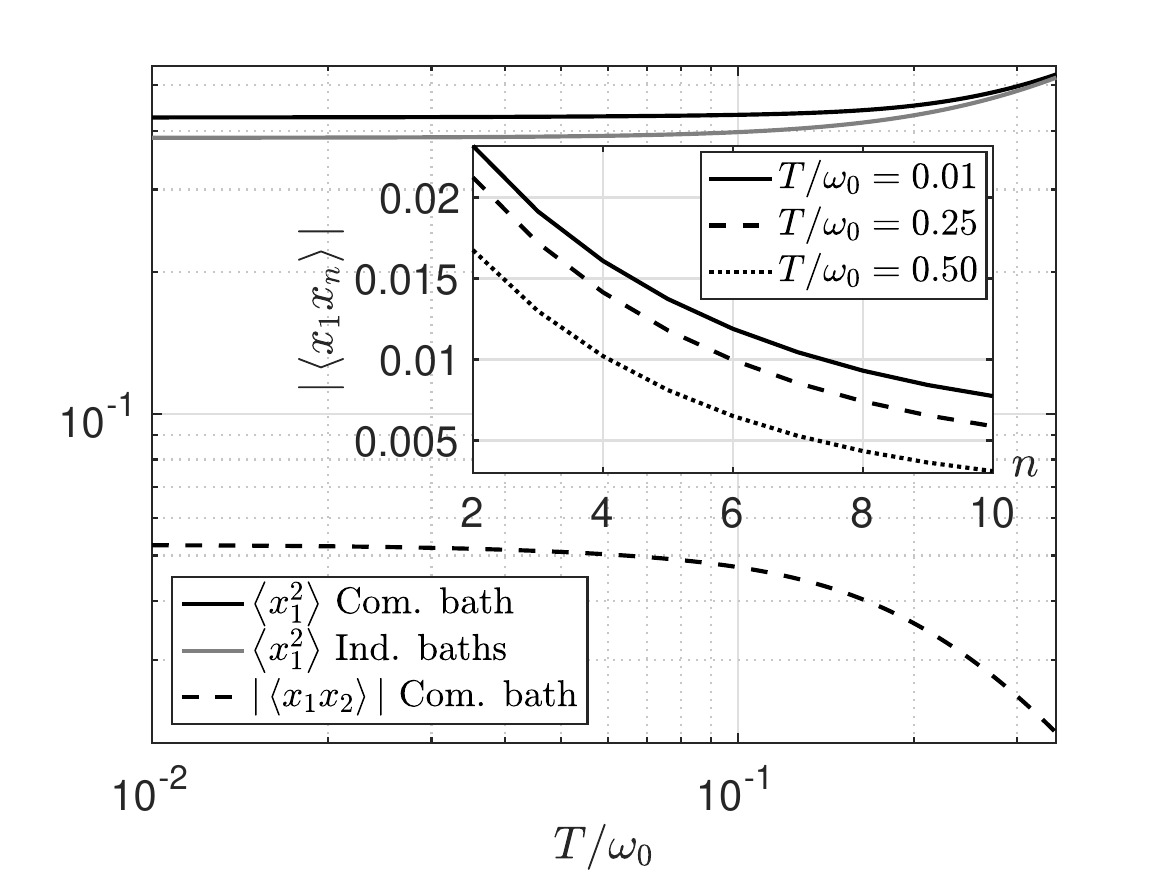}
		\includegraphics[width=.5\columnwidth]{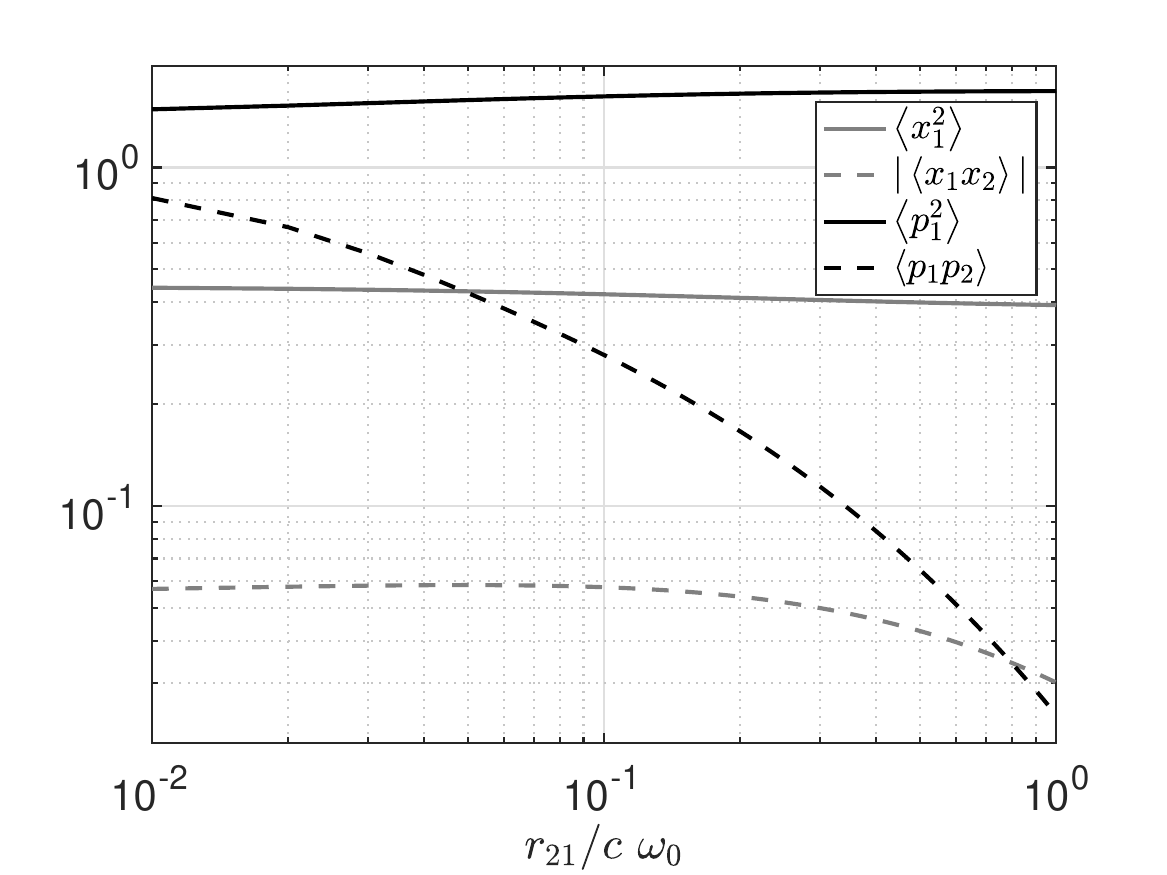}
	\end{tabular}
	\caption{Left.--- On site correlations as showcased by $\average{x_1^2}$ show slightly different behavior in the two scenarios; (a) in solid gray and (b) in solid black. The bath-induced inter-oscillator correlations like $|\average{x_1x_2}|$ only exist in the common bath scenario (b) (dashed black). These correlations vanish as temperature increases, and the two scenarios become equivalent. 
		\textbf{Inset}: Inter-oscillator correlation $|\average{x_1x_n}|$ in the common bath scenario, as a function of $n$ and for different temperatures; the farther the two oscillators are, the smaller their correlations are, as expected. The parameters are set to $N=10$, $\omega_0=1$, $\omega_i=\omega_0\forall i$, $\Omega=100\omega_0$, $\gamma=\sqrt{\omega_0}$, $g_{ij}=0$, and $r_{21}/c\omega_0=0.1$.
		Right.---Different correlations as a function of distance in the common bath scenario (b); $\average{x_1^2}$ in solid gray, $\average{p_1^2}$ in solid black, $|\average{x_1x_2}|$ in dashed gray, and $\average{p_1p_2}$ in dashed black. While the on site correlations slightly depend on the distance, the inter-oscillator correlations are affected by it more significantly. Again, as distance increases scenario (b) reduces to scenario (a). Here we set $N=10$, $\omega_0=1$, $\omega_i=\omega_0\forall i$, $\Omega=100\omega_0$, $\gamma=\sqrt{\omega_0}$, $g_{ij}=0$, and $T/\omega_0=0.01$. }\label{Fig:Correlations_vs_T}
\end{figure}
-------------------------------
-------------------------------
\begin{figure}
	\begin{tabular}{c}	\includegraphics[width=.5\columnwidth]{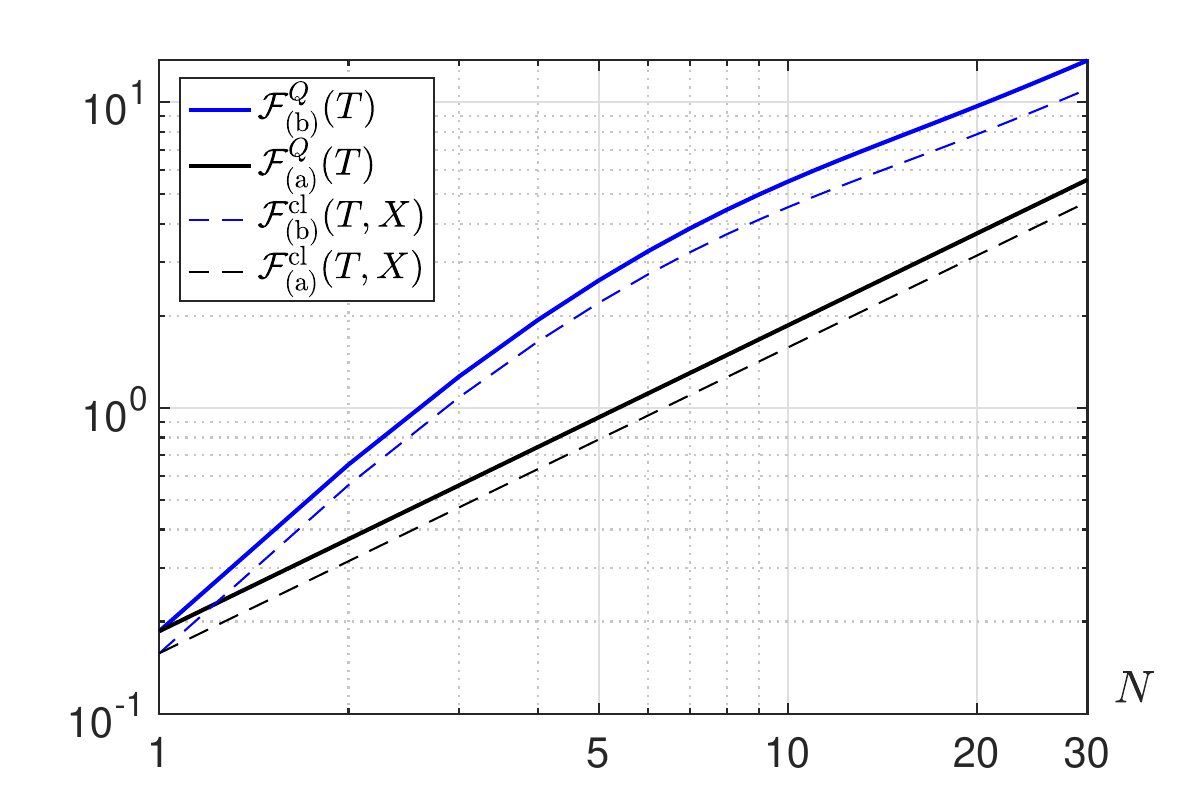}
		\includegraphics[width=.5\columnwidth]{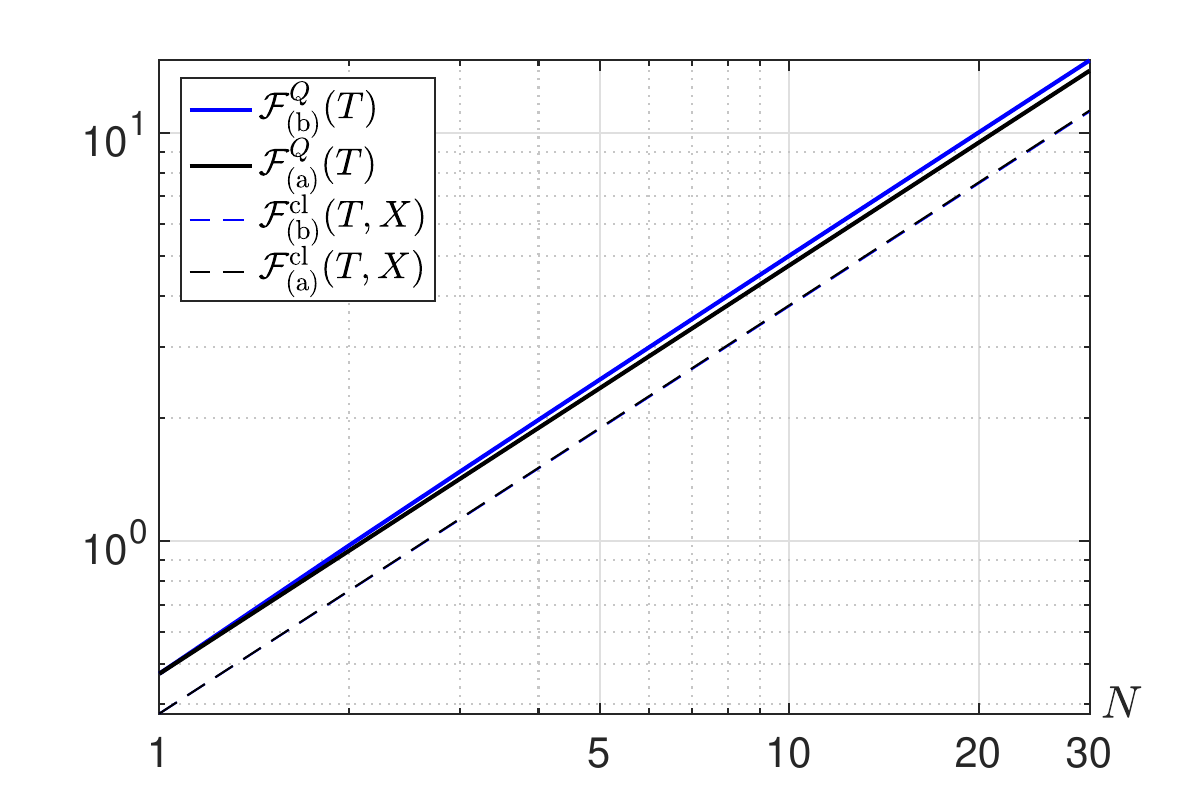}
	\end{tabular}
	\caption{Left.--- Loglog plot of the quantum and classical Fisher information vs $N$ for the common bath scenario (solid blue) and independent baths scenario (solid black). Here, we tune the setting to the \textit{intermediate temperature} limit with $T_0=0.1 \omega_0$. The common bath scenario shows a coefficient advantage in that the QFI in the common bath scenario is few times bigger than the QFI in the independent baths scenario. 
		Right.--- Similar to left, but with $T/\omega_0=1$. The two scenarios are in principle indistinguishable.	The rest of the parameters for both panels are set to $\omega_0=1$, $\omega_i=\omega_0\forall i$, $\gamma= \sqrt{\omega_0}$, $\Omega=100 \omega_0$, $g_{ij}=0$ and $r_{21}/c\omega_0=0.1$.}\label{Fig:QFI_scaling_int_T}
\end{figure}
-------------------------------
-------------------------------

In order to compute the steady state solution of our model, we comment on the spectral density $J_{ij}(\omega)$ and the dissipation kernel $\chi(\omega)$. 
To begin with, we consider an Ohmic form for the diagonal elements of the spectral density
\begin{align}
	J_{ii}(\omega)=\gamma^2 \omega \frac{\Omega^2}{\omega^2+ \Omega^2},
\end{align}
with $\gamma$ representing the strength of the interaction, and $\Omega$ being the cutoff frequency. We still have to find the off-diagonal terms.
By recalling the definition of the spectral density
\begin{align}
	\int_0^\infty J_{ij} (\omega)d\omega=\frac{\pi}{2 }\sum_{{k }} \frac{G^2_{ k}}{m_{{ k}} \omega_{ k}}\cos ( { k}  {r}_{ij} ) ,
\end{align}
and assuming a linear dispersion relation ${ k}=\omega_{ k}/c$ for our one-dimensional bath, we should have
\begin{align}\label{eq:SD_Freq}
	J_{ij}(\omega)={\gamma^2} \omega \frac{\Omega^2}{\omega^2+ \Omega^2}\cos \frac{\omega \left|{r}_{ij}\right|}{c},
\end{align}
that completely characterizes our spectral density.

Using the spectral density \eqref{eq:SD_Freq} one may find both the imaginary and the real parts of the susceptibility. Starting with the definition of $\chi(t)$, Eq.~(10), and using the isotropy of the bath gives
\begin{align}
	\chi_{ij}(t) = \sum_{ k} \frac{G^2_{ k}}{m_{ k}\omega_{ k}}\sin(\omega_{ k}t + { k}  { r}_{ij} )\Theta(t)
	=\sum_{ k} \frac{G^2_{ k}}{m_{ k}\omega_{ k}}\sin(\omega_{ k}t)\cos({ k}  { r}_{ij} )\Theta(t),
\end{align}
which by definition of the spectral density reads
\begin{align}
	\chi_{ij}(t) = \frac{2}{\pi}\int_{0}^{\infty}J_{ij}(\omega)\sin(\omega t)\Theta(t)d\omega=i\frac{ \Theta(t)}{\pi}\int_{-\infty}^{\infty}J_{ij}(\omega)e^{-i \omega t}d\omega=2i \Theta(t) F^{-1}\left( J_{ij}(\omega) \right) ,
\end{align}
with $F(\circ)$ [$F^{-1}(\circ)$] being the [inverse] Fourier transform of $(\circ)$, and we used the fact that $J_{ij}(\omega)$ is an odd function---making $J_{ij}(\omega)\cos \omega t$ and $J_{ij}(\omega)\sin \omega t$ odd and even, respectively. Taking the Fourier transform of this expression and using the fact that for our choice of normalization $F(f g)=\frac{1}{2 \pi} {\tilde f} \star {\tilde g}$ we get to
\begin{align}
	{\tilde \chi}_{ij}(\omega)=\frac{i}{\pi} \left[\frac{i}{\omega +i \varepsilon}\right]\star J_{ij}(\omega),
\end{align}
where $\varepsilon$ is an infinitesimal positive parameter. By defining $a_{ij} = |{ r}_{ij}|/c$, we can write
\begin{align}
	{\tilde \chi}_{ij}(\omega) & = \frac{-1}{\pi}\int d\omega^{\prime} \frac{J_{ij}( \omega^{\prime})}{(\omega-\omega')+ i \varepsilon} \\
	&= \frac{-{\gamma^2}\Omega^2}{2\pi}\int d\omega^{\prime}\frac{\omega^{\prime}e^{i\omega^{\prime}a_{ij}}}{(\omega-\omega' +i \varepsilon)(\omega^{\prime 2}+\Omega^2)}
	-\frac{{\gamma^2}\Omega^2}{2\pi}\int d\omega^{\prime}\frac{\omega^{\prime}e^{-i\omega^{\prime}a_{ij}}}{(\omega-\omega' +i \varepsilon)(\omega^{\prime 2}+\Omega^2)}\nonumber\\
	&= \frac{{\gamma^2}\Omega^2}{2\pi} \int d\omega^{\prime}(f(\omega^{\prime}) + g(\omega^{\prime})),
\end{align}
where we have defined $f(\omega^{\prime}) = \frac{\omega^{\prime}e^{i\omega^{\prime}a_{ij}}}{(\omega'-\omega-i\varepsilon)(\omega^{\prime 2}+\Omega^2)}$ and $g(\omega^{\prime}) = \frac{\omega^{\prime}e^{-i\omega^{\prime}a_{ij}}}{(\omega^{\prime}-\omega-i\epsilon)(\omega^{\prime 2}+\Omega^2)}$. 
Both integrands have three poles at $\omega^{\prime} \in \{\omega_0=\omega+i\epsilon,\omega_{\pm} = \pm i \Omega \}$. Then, the integrals can be calculated on the complex plane by choosing a circuit along the real axis closed by an arc on the upper or lower half-plane and counting the poles that lie inside the circuit. That is,
\begin{align}
	\int_{-\infty}^{\infty} d\omega^{\prime} f(\omega^{\prime}) &= 2 i \pi \sum {\rm Res}(\text{upper half plane})\\
	&=2 i \pi \left(\left[ (\omega^{\prime}-\omega_+) f(\omega^{\prime})\right]_{\omega^{\prime}=\omega_+} 
	+  \left[ (\omega^{\prime}-\omega_0) f(\omega^{\prime})\right]_{\omega^{\prime}=\omega_0}\right)
	\nonumber\\
	&=2 i\pi \left( \frac{i\Omega e^{- \Omega  a_{ij}}}{(i \Omega - \omega)(2 i \Omega)}+\frac{\omega e^{i \omega a_{ij}}}{(\omega^2+ \Omega^2)}\right),
\end{align}
and similarly,
\begin{align}
	\int_{-\infty}^{\infty} d\omega^{\prime} g(\omega^{\prime}) 
	&=-2 i\pi \sum {\rm Res}(\text{lower half plane})\\
	&=-2 i \pi \left[ (\omega^{\prime}-\omega_-) g(\omega^{\prime})\right]_{\omega^{\prime}=\omega_-} \nonumber\\
	& 
	= -2 i\pi \frac{-i\Omega e^{-\Omega a_{ij}}}{(-i\Omega -\omega)(-2i\Omega)}
	= \frac{i \pi e^{- \Omega a_{ij}}}{\omega+i\Omega}.
\end{align}
Note that we have to close the circuits as indicated so that the real part of the exponent in each corresponding exponential is negative and the integral over each arc does not contribute. This also means that in the case of $g(\omega')$ a sign has to be added because the circuit is now clockwise. Putting everything together we get
\begin{align}
	{\tilde \chi}_{ij}(\omega) & =  \frac{{\gamma^2}\Omega^2}{2\pi}\int d\omega^{\prime}(f(\omega^{\prime}) + g(\omega^{\prime})) =\frac{{\gamma^2}\Omega^2}{2\pi} \left( \frac{ i\pi e^{-\Omega a_{ij}}}{ i\Omega - \omega}+\frac{2 i \pi \omega e^{i \omega a_{ij}}}{\omega^2+\Omega^2}+\frac{ i \pi e^{-\Omega a_{ij}}}{\omega + i \Omega}\right)
	= \frac{{\gamma^2} \Omega^2}{\omega^2+ \Omega^2} \left(\Omega e^{-\Omega a_{ij}}+i\omega e^{i \omega a_{ij}}  \right)
\end{align}

{\it Renormalization of the Hamiltonian.---}
The probe-bath interaction affects the effective Hamiltonian of the probe, which is additional to the dissipative role of the bath. As a result, at large temperatures, the probe will thermalise to a Gibbs state with its effective Hamiltonian, rather than the physical one. In order to only focus on the dissipation impact of the bath, it is common practice to add a counter term to renormalise the Hamiltonian and hence have a thermal state with system's original Hamiltonian~\cite{weiss2012quantum,breuer2002theory,PhysRevA.96.062103}. 
%
The position and momentum of bath oscillators at minimum is given by
\begin{align}
	\left.\frac{\partial H}{\partial y_{{k}}}\right|_{y_{{k}}^*,q_{{k}}^*}=\left.\frac{\partial H}{\partial q_{{k}}}\right|_{y_{{k}}^*,q_{{k}}^*}=0\quad \Longrightarrow\quad y^*_{ {k}}=-\sum_i\frac{ G_{{k}}}{m_{{k}} \omega^2 _{{k}}} x_i\cos( {k}  {r}_i),\qquad q^*_{ {k}}=-\sum_i\frac{ G_{{k}}}{ \omega _{{k}}} x_i\sin( {k}  {r}_i),
\end{align}
as follows immediately by taking partial derivatives of the Hamiltonian (1). Substituting in the total Hamiltonian one gets
\begin{align}
	H_{\rm eff}({y_{{k}}^*,q_{{k}}^*})&=H_P+ \left.H_B\right|_{y_{{k}}^*,q_{{k}}^*} + \left.H_{PB}\right|_{y_{{k}}^*,q_{{k}}^*}\nonumber\\
	&=\sum_i\frac{p_i^2}{2 m_i}+ \frac{1}{2}\sum_i m_i \omega_i^2 x_i^2+\frac{1}{2}\sum_{j\neq i}g_{ij} x_i x_j -\frac{1}{2}\sum_{{k}, i,j} \frac{G_{{k}}^2}{m_{{k}} \omega_{{k}}^2}x_i x_j\cos({k}  { r}_{ij}).
\end{align}
This corresponds to the effective Hamiltonian of the system when disregarding fluctuations of the bath around the minimum, which is shifted from its original Hamiltonian. This can be brought back to the original form in the absence of the bath i.e., Eq.~(2), by adding the following counter-term to the total Hamiltonian
\begin{align}
	\Delta H
	=\frac{1}{2} \sum_{{k},i,j} \frac{G^2_{{k}}}{ m_{{k}} \omega^2_{{k}} }
	\cos({k}  { r}_{ij}) x_i x_j
	=\frac{1}{\pi}\sum_{i,j} x_i x_j\int_0^\infty \frac{J_{ij}(\omega)}{\omega} d\omega.
\end{align}
For our model this integral can be computed analytically
\begin{align}
	\frac{2}{\pi}\int_0^\infty \frac{J_{ij}(\omega)}{\omega}d \omega&= \frac{2}{\pi}\int_0^\infty \frac{{\gamma^2} \Omega^2}{\omega^2+ \Omega^2}\cos (\omega a_{ij})= {\gamma^2} \Omega e ^{- \Omega a_{ij}}.
\end{align}
Finally, by defining $g_{ii}\coloneqq m_i \omega_i^2$, the counter-term can be absorbed as a shift in the frequency and the couplings
\begin{align}\label{app:renormalization_couplings}
	g_{ij}\to g_{ij}+{\gamma^2} \Omega e ^{- \Omega a_{ij}} \eqqcolon g^R_{ij}
\end{align}
{\it The temperature and distance profile of the correlations.---}\label{app:Cov_Mat}
As we mentioned in the main text, the inter-oscillator correlations vanish at both high $T$ and long distance limits, and the two scenarios (a) and (b) become equivalent. We showcase this in Fig.~\ref{Fig:Correlations_vs_T}. In the left panel we depict the on-site $\average{x_1^2}$ correlation for both scenarios (a) and (b), as well as inter-oscillator correlations $\average{x_1x_2}$ that are only present in scenario (b), against temperature. Although at low temperatures (a) and (b) have different covariance matrices, as $T$ increases the on-site correlations behave the same, while the inter-oscillator correlations in (b) disappear. 
In the inset we picture the behavior of the inter-oscillator correlations $\average{x_1x_n}$ against ``$n$" for different temperatures. It is seen that correlations are present not just among next neighbors, but also for farther distances in the chain. As one considers more distant oscillators the correlations between them decreases, as expected. Similar behaviours are seen for the momentum operators (not plotted here).

In the right panel of Fig.~\ref{Fig:Correlations_vs_T} we plot the on-site correlations $\average{x_1^2}$ and  $\average{p_1^2}$, as well as next-neighbor correlations $|\average{x_1x_2}|$ and $\average{p_1p_2}$, as a function of the distance between next-neighbors in the chain $r_{21}$. While on-site correlations remain relatively less affected by increasing the distance, the inter-oscillator correlations disappear quickly, and thus the scenario (b) converges into scenario (a) at the large $r_{ij}$ limit. Again, similar behaviour is seen for other elements of the covariance matrix.
\begin{figure}
	\includegraphics[width=.49\linewidth]{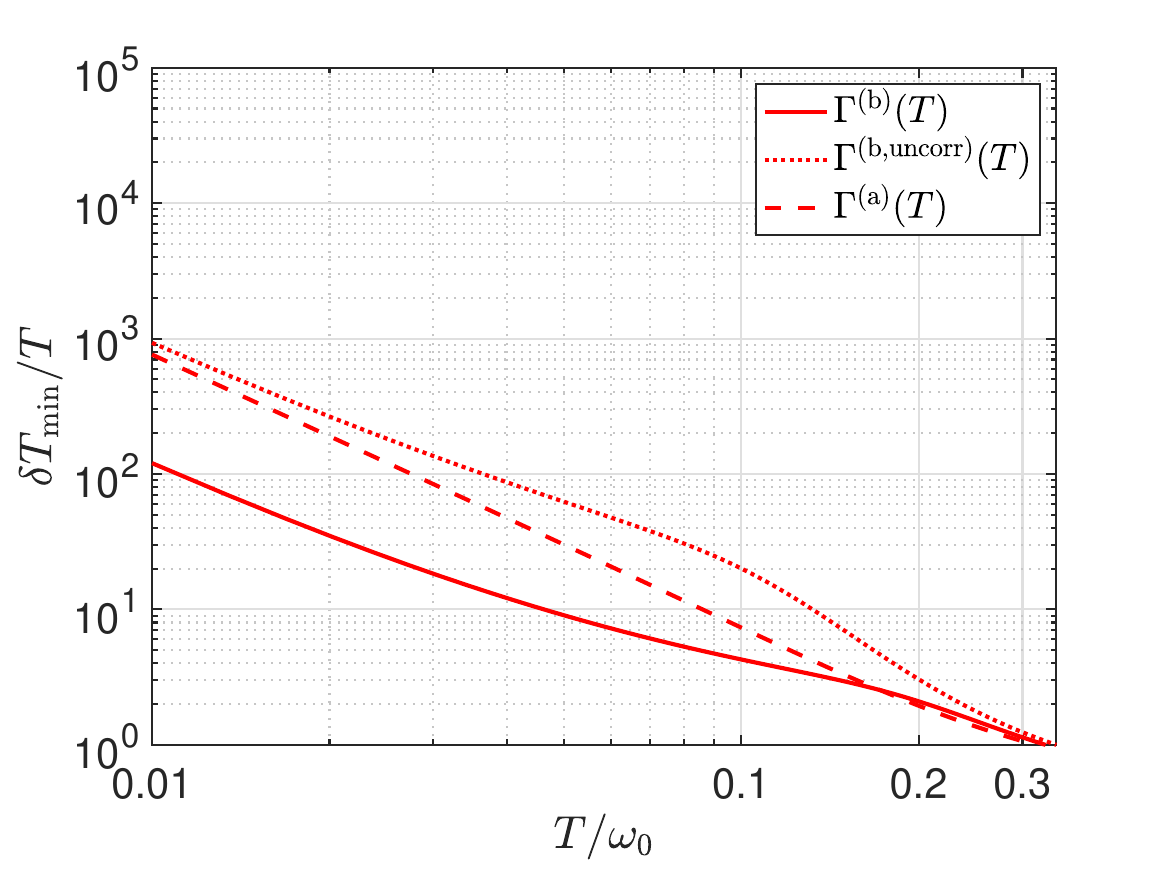}
	\includegraphics[width=.49\linewidth]{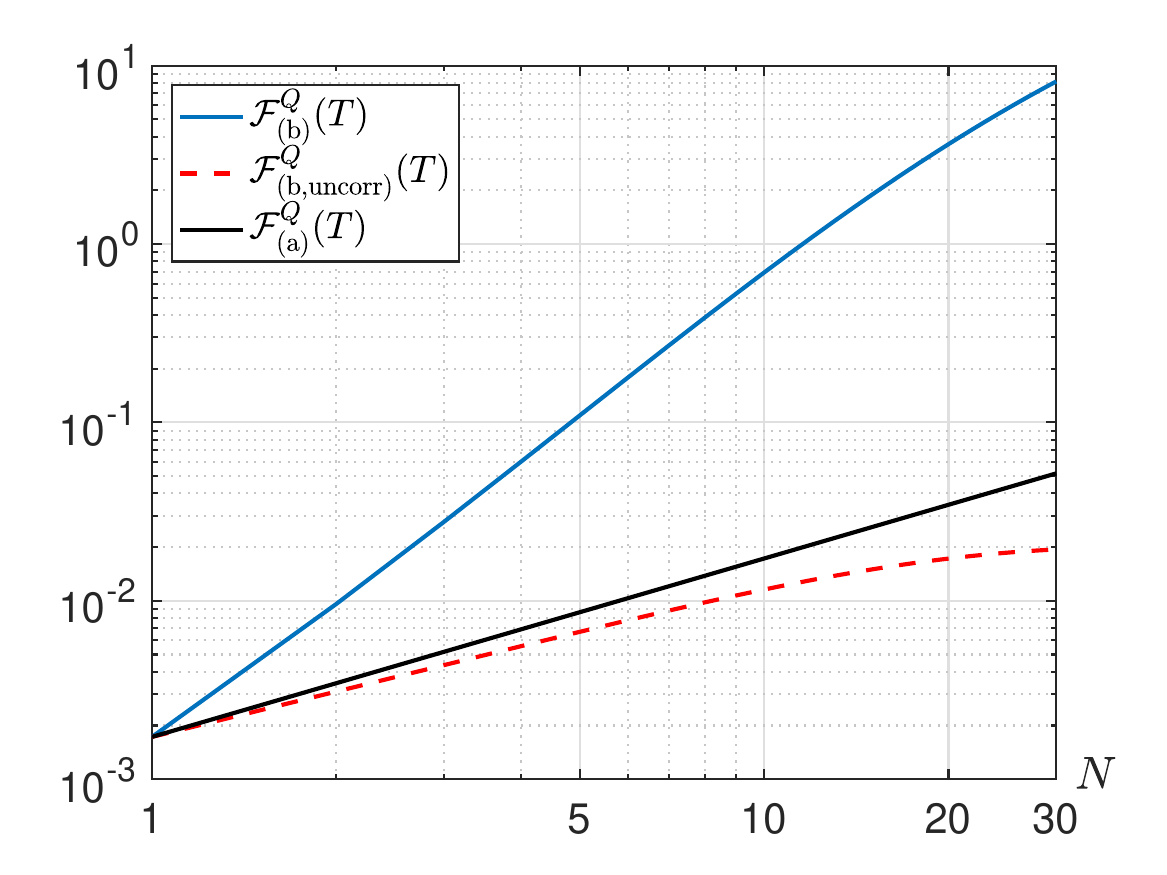}
	\caption{Left---Similar to Fig. 2 of the main text, only for $\gamma=\sqrt{\omega_0}$. The new dotted line introduced here corresponds to the common baths scenario but without the correlations, so that one can see how the common bath scenario performs if we ignore the impact of the correlations. Right---Similar to Fig. 4 of the main text (only the QFI is depicted). Again, for comparison we have introduced the red dashed line corresponding to the common bath scenario but without correlations. It is clear that correlations are the major reason for the enhancement in thermometry.}\label{fig:corr_uncorr}
\end{figure}
While the inter-oscillator correlations that we see here are the main difference between scenario (b) and scenario (a) they are not the only one. The common bath can in general have two kinds of impacts. Let us denote with $\sigma_{i}^{(\rm \alpha)}(T)$ the local covariance matrix of the $i$th probe oscillator in scenario $\alpha \in \{\rm a,b\}$. The two differences are the followings: (i) In general we do not have $\sigma_i^{(\rm a)} = \sigma_i^{(\rm b)}$ (i.e., at the level of individual oscillators the two scenarios may look different), and (ii) In the common bath scenario (b) inter-oscillator correlations exist, i.e., $\Gamma^{(b)}(T)\neq \oplus_i \sigma_i^{(\rm b)}(T) $ while in contrast we have $\Gamma^{(a)}(T) = \oplus_i \sigma_i^{(\rm a)}(T) $ for the independent baths scenario. A way to confirm that the correlations are responsible for enhanced thermometry precision is to consider a third covariance matrix, that is $\Gamma^{(\rm b, uncorr)}(T) \coloneqq \oplus_i \sigma_i^{(\rm b)}(T) $  i.e., the covariance matrix of the common-bath scenario without all the correlations. In Fig.~\ref{fig:corr_uncorr} we look at the quantum Fisher information, and the relative error for this covariance matrix and compare it with the scenarios (a) and (b). We can see that, indeed, without the correlations $\Gamma^{(\rm b, uncorr)}(T)$ performs even worse than the independent baths scenario (a). Thus, one can conclude that the correlations are responsible for the enhancement in precision. 
We emphasise that these kind of bath-induced correlations might be but are not necessarily quantum correlations. 
In fact, for our specific model and the parameters that we considered, no entanglement as measured by the entanglement negativity for any arbitrary bipartition is present, while the mutual information is always non-zero, showing that the enhancement in thermometric precision is due to classical correlations among probes~\cite{weedbrook2012gaussian}\footnote{Despite discord being only non-vanishing for product Gaussian states~\cite{PhysRevA.87.012119}, It would be interesting to check the behaviour its in our setting, however, in our specific problem discord is not directly calculable, as it requires an optimisation problem \cite{PhysRevLett.105.020503,adesso2011gaussian}.}. 

\section{Parameter estimation in Gaussian quantum systems}
In this paper we have used results from quantum metrolgy in order to find the ultimate bounds on thermometry of our Bosonic sample. Here, we remind the main techniques that were originally proven in~\cite{monras2013phase}.
Firstly, the symplectic form can be defined in terms of the canonical commutation relations
\begin{align}
	[R_i,~R_j] = i\omega_{ij}, \hspace{1cm} \Omega = -\omega. 
\end{align}
Gaussian states are fully determined by their first and second moments
\begin{align}
	d = {\rm tr}[\rho R], \hspace{1cm} \Gamma = {\rm tr} [(R-d)\circ(R-d)^T \rho],
\end{align}
where $d\in {\mathbb R}^{2N}$, and $\Gamma \in {\cal M}_{2N} (\mathbb{R})$. We also defined the symmetric product as $A\circ B = \frac{1}{2}(AB + BA)$.
\\ 
{\it The optimal measurement.---}\label{App:SLD}
The optimal measurement is a projective one carried out in the basis of the symmetric logarithmic derivative (SLD). The latter is a self adjoint operator $\Lambda$ that satisfies the following equation
\begin{align}\label{eq:SLD_def}
	\Lambda\circ \rho = \partial_{\theta} \rho,
\end{align}
with $\theta$ being the parameter to be estimated.
One can prove that the SLD is at most 2nd order in the quadrature operators and reads
\begin{align}\label{eq:SLD}
	\Lambda = L^{(0)} + L_i^{(1)} R_i + L_{ij}^{(2)} R_i \circ R_j,
\end{align}
where we use Einstein's summation rule. The matrix $L^{(2)}$ is the solution of the following matrix equation
\begin{align}
	\partial_{\theta} \Gamma =  2\Gamma L^{(2)} \Gamma+\frac{1}{2}\Omega L^{(2)} \Omega ,
\end{align}
which can be straightforwardly solved by vectorization. Using $L^{(2)}$ one can find the vector $L^{(1)}$ through
\begin{align}
	L^{(1)} = \Gamma^{-1}\partial d - 2L^{(2)} d,
\end{align}
which vanishes if $d=0$. Finally, the constant $L^{(0)}$ is given by
\begin{align}
	L^{(0)} = -L^{(1)T}d - \frac{1}{2}{\rm tr}[L^{(2)} \Gamma] - d^T L^{(2)} d
\end{align}
\\
{\it The quantum Fisher information.---}\label{App:QFI}The
quantum Fisher information is defined as
\begin{align}
	{\cal F}(\rho) & = {\rm tr}[\rho \Lambda^2] = {\rm tr}[\Lambda\rho\circ\Lambda],
\end{align} 
which by using \eqref{eq:SLD_def} and \eqref{eq:SLD} reads
\begin{align}
	{\cal F}(\rho) 
	& = {\rm tr}\left[\left\{L^{(0)} + L_i^{(1)} R_i + L_{ij}^{(2)} R_i \circ R_j\right\}\partial\rho\right]
	\nonumber\\
	& = L^{(1)T}\partial d + L^{(2)}_{ij}\partial{\rm tr} [(R_i\circ R_j)  \rho]\nonumber\\
	& = 
	\partial d^T \Gamma^{-1} \partial d - 2{\rm Tr}[ L^{(2)} (\partial d)d^T] + L^{(2)}_{ij}\partial{\rm tr} [(R_i\circ R_j)  \rho]\\
	& = 
	\partial d^T \Gamma^{-1} \partial d + {\rm Tr}[L^{(2)} \partial\Gamma]\nonumber\\
	&= \partial d^T \Gamma^{-1} \partial d +\frac{1}{2} \bra{\partial \Gamma} (\Gamma\otimes\Gamma -\frac{1}{4} \Omega\otimes\Omega)^{-1} \ket{\partial \Gamma},\\
	& = \partial d^T \Gamma^{-1} \partial d +2 {\rm Tr}[L^{(2)}\Gamma L^{(2)}\Gamma + \frac{1}{4}L^{(2)}\Omega L^{(2)}\Omega],
\end{align}
where we define ${\rm Tr}$ for the space of covariance matrices, while ${\rm tr}$ operates on the Hilbert space of the density matrices, and also used a vectorization notation in which an $m\times m$ matrix $A$ is vectorized to an $m^2$ column with the rule $ \ket{A}_{m(i-1)+j} = A_{ij}$.
This last equation provides the QFI for a Gaussian system. 
{\it }
According to \cite{monras2013phase}, if we consider proper dimensions, i.e., if we consider $\Omega \mapsto \hbar \Omega$ and let $\hbar \to 0$, we revive the classical Fisher information for a \textit{classical Gaussian probability distribution} 
\begin{align}\label{eq:FI_Classical}
	{\cal F}_{\rm cl} = \partial d^T \Gamma^{-1} \partial d + 2{\rm Tr}[L^{(2)}\Gamma L^{(2)}\Gamma] = \frac{1}{2} \bra{\partial \Gamma} \Gamma^{-1}\otimes\Gamma^{-1}\ket{\partial \Gamma} = \frac{1}{2}{\rm Tr}[\Gamma^{-1}(\partial \Gamma) \Gamma^{-1}(\partial \Gamma)].
\end{align}
This latter equation is directly proven for classical systems in \cite{malago2015information} and revived in \cite{Nina_et_al_in_preparation}. 
This is specifically useful when we perform sub-optimal Gaussian measurements on a quantum Gaussian system. Suppose we perform a Gaussian measurement $\Pi^{s}$, which can be represented by a physical covariance matrix $\Gamma^s \geq i\Omega$, while the Gaussian system to be measured has its own covariance matrix $\Gamma \geq i\Omega$. Then, the outputs of this measurement are Gaussian distributed with the classical covariance matrix ${\tilde \Gamma} = \Gamma^s + \Gamma \geq 0$. The classical Fisher information associated to this Gaussian measurement is given by \eqref{eq:FI_Classical} by replacing $\Gamma \to {\tilde \Gamma}$.
In this work, we studied the behaviour of local position measurement $X=\otimes_{i=1}^N x_i$. The covariance matrix associated to this measurement is given by 
$\Gamma^X = \lim_{R\to\infty} \oplus_{i=1}^N {\rm diag}[1/R, R]$.
For the local momentum measurement $P=\otimes_{i=1}^N p_i$ a similar calculation was done by using $\Gamma^P = \lim_{R\to\infty} \oplus_{i=1}^N {\rm diag}[R, 1/R]$, but since its precision is not as good as the position measurement, we did not present the plots here. 

\end{document}